\documentclass[apj]{emulateapj}
\usepackage{amsmath,graphicx,epsfig,multirow,natbib,rotating,tabularx}
\usepackage[usenames]{color}
\shorttitle{Joint Lensing Analysis of J0850}
\shortauthors{WONG et al.}

\begin{document}

\newcommand{\comment}[1]{\textcolor{red}{\textbf{#1}}}
\newcommand{\zl}{0.3774}
\newcommand{\zfg}{0.2713}
\newcommand{\mclwl}{3.15_{-0.81}^{+1.13} \times 10^{15}~\mathrm{M_{\odot}}}
\newcommand{\mclsl}{2.93_{-0.65}^{+0.71} \times 10^{15}~\mathrm{M_{\odot}}}
\newcommand{\mfgwl}{6.22_{-3.62}^{+9.21} \times 10^{13}~\mathrm{M_{\odot}}}
\newcommand{\mfgsl}{6.53_{-4.02}^{+9.24} \times 10^{13}~\mathrm{M_{\odot}}}
\newcommand{\csl}{3.46_{-0.59}^{+0.70}}

\title{Joint Strong and Weak Lensing Analysis of the Massive Cluster Field J0850+3604}
\author{
Kenneth C. Wong\altaffilmark{1,2,6},
Catie Raney\altaffilmark{3},
Charles R. Keeton\altaffilmark{3},
Keiichi Umetsu\altaffilmark{2},
Ann I. Zabludoff\altaffilmark{4},
S. Mark Ammons\altaffilmark{5},
and K. Decker French\altaffilmark{4}}
\altaffiltext{1}{National Astronomical Observatory of Japan, Mitaka, Tokyo 181-8588, Japan}
\altaffiltext{2}{Institute of Astronomy and Astrophysics, Academia Sinica (ASIAA), P.O. Box 23-141, Taipei 10617, Taiwan}
\altaffiltext{3}{Department of Physics and Astronomy, Rutgers University, 136 Frelinghuysen Road, Piscataway, NJ 08854}
\altaffiltext{4}{Steward Observatory, University of Arizona, 933 North Cherry Avenue, Tucson, AZ 85721}
\altaffiltext{5}{Lawrence Livermore National Laboratory, 7000 East Avenue, Livermore, CA 94550}
\altaffiltext{6}{EACOA Fellow}

\begin{abstract}
We present a combined strong and weak lensing analysis of the J085007.6+360428 (J0850) field, which was selected by its high projected concentration of luminous red galaxies and contains the massive cluster Zwicky 1953.  Using Subaru/Suprime-Cam $BVR_{c}I_{c}i^{\prime}z^{\prime}$ imaging and MMT/Hectospec spectroscopy, we first perform a weak lensing shear analysis to constrain the mass distribution in this field, including the cluster at $z = 0.3774$ and a smaller foreground halo at $z = 0.2713$.  We then add a strong lensing constraint from a multiply-imaged galaxy in the imaging data with a photometric redshift of $z \approx 5.03$.  Unlike previous cluster-scale lens analyses, our technique accounts for the full three-dimensional mass structure in the beam, including galaxies along the line of sight.  In contrast with past cluster analyses that use only lensed image positions as constraints, we use the full surface brightness distribution of the images.  This method predicts that the source galaxy crosses a lensing caustic such that one image is a highly-magnified ``fold arc", which could be used to probe the source galaxy's structure at ultra-high spatial resolution ($< 30$ pc).  We calculate the mass of the primary cluster to be $\mathrm{M_{vir}} = 2.93_{-0.65}^{+0.71} \times 10^{15}~\mathrm{M_{\odot}}$ with a concentration of $\mathrm{c_{vir}} = 3.46_{-0.59}^{+0.70}$, consistent with the mass-concentration relation of massive clusters at a similar redshift.  The large mass of this cluster makes J0850 an excellent field for leveraging lensing magnification to search for high-redshift galaxies, competitive with and complementary to that of well-studied clusters such as the {\it HST} Frontier Fields.
\end{abstract}


\section{Introduction} \label{sec:intro}
Gravitational lensing by galaxy clusters is a powerful tool to study faint background objects through lensing magnification.  These massive objects can strongly lens background galaxies into multiple images or giant arcs on scales arcseconds to tens of arcseconds \citep[e.g.,][]{fort1994,bartelmann1998,kneib2011}, as well as weakly lens many more background galaxies across scales of arcminutes by inducing small tangential distortions in their observed shapes \citep[e.g.,][]{kaiser1993,bartelmann2001,hoekstra2013}.  By taking advantage of the magnification provided by these cosmic telescopes, it is possible to observe and study the properties of the earliest, most distant galaxies at $z \gtrsim 7$.  Many orbits of {\it Hubble Space Telescope} ({\it HST}) time have been dedicated to characterizing these clusters and searching for lensed high-$z$ galaxies to better understand their properties and determine their contribution to the reionization of the intergalactic medium during this epoch.

The {\it HST} Frontier Fields program \citep[HFF;][]{lotz2017} is imaging six of the most massive and well-studied lensing clusters with 140 orbits of Advanced Camera for Surveys (ACS) and Wide-Field Camera 3 (WFC3) IR channel observations.  This program is detecting and constraining the properties of the first generation of galaxies at $z \gtrsim 7$, particularly those at the faint end of the galaxy luminosity function that are otherwise inaccessible without the aid of lensing magnification.  These observations have also identified new multiply-imaged systems at lower redshifts.  These constraints improve the cluster mass models and magnification maps, which are crucial for understanding the intrinsic properties of the high-$z$ galaxies detected in these fields.

While high-$z$ galaxies have been identified in the HFF \citep[e.g.,][]{zitrin2014,atek2015a,atek2015b,coe2015,infante2015,ishigaki2015,ishigaki2017,kawamata2016,laporte2016}, those fields have not produced the expected number of detections \citep[e.g.,][]{coe2015,laporte2016}, suggesting either faster-than-expected evolution in the galaxy luminosity function at high-$z$, or systematic uncertainties leading to this result.  Even with six independent fields, cosmic variance is expected to be a significant source of uncertainty \citep{robertson2014,bouwens2015}.  For example, \citet{ishigaki2017} find that the evolution of the cosmic UV luminosity density is consistent with a smooth linear evolution with redshift, citing cosmic variance as a potential reason for their disagreement with their previous results in \citet{ishigaki2015}.  Exploring new lines of sight for future use as cosmic telescopes will be key to addressing the cosmic variance issue as we move into the era of the {\it James Webb Space Telescope} ({\it JWST}).

\citet{wong2013} identified 200 fields in the Sloan Digital Sky Survey (SDSS) data release 9 \citep[DR9;][]{ahn2012} containing the largest integrated luminosity in luminous red galaxies (LRGs).  LRGs are tracers of massive group and cluster-scale structures, indicating that these fields may contain the most massive lensing clusters, and in some cases, multiple clusters projected along the line of sight.  Such configurations can increase the lensing cross section, making them better probes of high-$z$ galaxies than single clusters of equal total mass \citep{wong2012,french2014}.  One of the best fields as ranked by this metric contains Abell 370, the only HFF target contained within the \citet{wong2013} survey volume, demonstrating the effectiveness of this selection technique.  Further exploration and characterization of these fields will identify the most promising gravitational lensing configurations and improve the magnification maps, setting the stage for future deep observations to detect and study the earliest generation of galaxies, complementary to the HFF clusters.

We are conducting an ongoing observational campaign using deep photometric and spectroscopic data to characterize the mass distribution in these unique fields.  An initial analysis of two of these fields confirms the presence of multiple cluster-scale halos and suggests total integrated masses of $\sim3\times10^{15} \mathrm{M_{\odot}}$, making them among the most massive lines of sight known \citep{ammons2014}.  The imaging data also reveal a handful of lensed arcs, which can be used to constrain strong lensing models of these systems.  Crude mass models of these fields based on dynamical masses are presented in \citet{ammons2014}, although there are large uncertainties due to the lack of information on key properties of the cluster halos such as concentration, ellipticity, and orientation.

Additional constraints from gravitational lensing are needed to understand the detailed mass distribution and determine magnification maps in these fields.  Both strong and weak lensing provide complementary information on the mass profile of massive galaxy clusters.  Deep space-based imaging is ideal for identifying multiply-imaged sources for a strong lensing analysis or background galaxies over a wide area for a weak lensing analysis, but obtaining such data for a large sample of clusters is observationally expensive.  If reasonable lens models can be constructed from ground-based data, it would be far more feasible to explore these fields, characterize the physical properties of the clusters in them, and leveraging their magnification properties for studying distant galaxies.  For fields in which there are multiple structures at distinct redshifts, a full treatment of the multi-plane lensing effects is needed.  If the most massive lines of sight are in fact likely to contain multiple structures in projection \citep[e.g.,][]{bayliss2014,french2014}, a consistent framework for dealing with these effects is needed to leverage those fields for studying the high-redshift universe.

In this paper, we present a combined strong-and-weak lensing analysis of the J085007.6+360428 (hereafter J0850) field, from which we can constrain its mass model and magnification properties.  This field was identified by \citet{wong2013} as one of the 200 most massive lines of sight in the SDSS.  Based on our follow-up spectroscopic observations of a subset of these fields \citep{ammons2014} and available archival multiband imaging, J0850 stands out as one of the most promising cosmic telescopes due to visually-identified lensed arcs and the presence of a massive galaxy cluster, Zwicky 1953 \citep{zwicky1961}, at a redshift of $z_{\mathrm{L}} = \zl$.  \citet{hao2010} also identified two Gaussian Mixture Brightest Cluster Galaxy (GMBCG) associations within $3\farcm5$ of the field center.  The X-ray temperature of the main cluster is $\langle kT \rangle = 14.5$ keV from {\it ROSAT} \citep{ebeling1998} and 7.37 keV from {\it Chandra} \citep{cavagnolo2009}.

This paper is organized as follows.  In Section~\ref{sec:data}, we summarize the photometric and spectroscopic data used in this analysis.  Our methodology for constraining the mass model of the cluster using weak and strong lensing constraints is described in Section~\ref{sec:lensing}.  We present our results in Section~\ref{sec:results} and summarize our main conclusions in Section~\ref{sec:conclusion}.  We assume $\Omega_{m} = 0.274$, $\Omega_{\Lambda}=0.726$, and $H_{0} = 71$\,km\,s$^{-1}$\,Mpc$^{-1}$.  All quantities are given in $h_{71}$ units unless otherwise indicated.  At $z_{\mathrm{L}} = \zl$, the angular scale is $1\arcmin \approx 309$ kpc.  All magnitudes given are on the AB system.

\section{Data} \label{sec:data}
The imaging and spectroscopic data used in this analysis are presented in \citet{ammons2014}.  We provide a summary here.

\subsection{Imaging Data} \label{subsec:imaging}
Subaru/Suprime-Cam \citep{miyazaki2002} imaging data for the J0850 field in $BVR_{c}I_{c}i^{\prime}z^{\prime}$ bands were obtained from the Subaru-Mitaka-Okayama-Kiso Archive \citep[SMOKA;][]{baba2002}.  The data were taken as a part of the Massive Cluster Survey \citep[MACS;][]{ebeling2001} follow-up program and were first published by \citet{hashimoto2008}.  The Suprime-Cam data are reduced using the SDFRED1 package \citep{yagi2002,ouchi2004}, and the astrometric solution is derived by matching to sources in SDSS using the astrometry software developed by \citet{lang2010}.  The details of the photometric data are in Table~\ref{tab:phot}.

\renewcommand*\arraystretch{1.5}
\begin{table}
\caption{Subaru Suprime-Cam Photometry for J0850 \label{tab:phot}}
\begin{ruledtabular}
\begin{tabular}{cccc}
Filter &
Observation Date(s) &
Depth\tablenotemark{a} &
Exp. Time (min)
\\
\tableline
$B$ &
2006 Dec 20 &
27.3 &
44
\\
$V$ &
2004 Feb 23; 2005 Nov 29 &
27.3 &
52
\\
$R_{c}$ &
2000 Dec 26; 2005 Mar 4$-$5 &
27.5 &
70
\\
$I_{c}$ &
2006 Dec 26 &
26.7 &
56
\\
$i^{\prime}$ &
2005 Mar 5 &
26.6 &
30
\\
$z^{\prime}$ &
2003 Apr 26; 2005 Mar 5 &
26.2 &
62
\\
\end{tabular}
\end{ruledtabular}
\tablenotetext{1}{3$\sigma$ sensitivities are calculated from final stacked images using 1\farcs5 diameter apertures.}
\end{table}
\renewcommand*\arraystretch{1.0}

\subsection{Spectroscopic Data} \label{subsec:spec}
We obtained spectra of 627 galaxies in the J0850 field with Hectospec \citep{fabricant2005,mink2007}, a multi-object fiber spectrograph on the MMT telescope.  The data are reduced using HSRED\footnote{http://code.google.com/p/hsred/}, a modification of the IDL SDSS pipeline written by R. Cool \citep{papovich2006}.  The targets are selected based on a $i < 21.1$ mag cut and the SDSS morphological star/galaxy discriminator.  Galaxies within $7\arcmin$ of the field center are prioritized, with lower priority given to galaxies out to $15\arcmin$ from the field center.

\section{Lensing Analysis} \label{sec:lensing}
In this section, we describe our methodology for using our imaging and spectroscopic data to constrain the properties of the J0850 field using both strong and weak lensing information.  Our general procedure is to first use the spectroscopic identification of cluster-scale halos along the LOS from the earlier analysis of \citet{ammons2014} as priors for our weak lensing analysis of the Subaru/Suprime-Cam data.  The posterior parameter distributions from the weak lensing analysis are then used as priors to generate Markov Chain Monte Carlo (MCMC) realizations of the full line-of-sight (LOS) mass distribution.  These models are then processed by the {\sc lensmodel}+{\sc pixsrc} software package \citep{keeton2001,tagore2014} to apply our strong lensing constraints and generate the final posterior parameter distribution.  This methodology is completely general, and can be applied to lines of sight containing multiple cluster-scale halos in projection.

\subsection{Constraints from Imaging and Spectroscopy}
An initial mass model of the J0850 field based on dynamical mass estimates is presented in \citet{ammons2014}.  Their analysis confirms the presence of a massive ($\mathrm{M_{vir}} = 3.2 \pm 0.3 \times10^{15}~\mathrm{M}_{\odot}$) cluster at $z_{\mathrm{L}} = \zl$, as well as a smaller foreground group ($\mathrm{M_{vir}^{fg}} = 6 \pm 4 \times10^{13}~\mathrm{M}_{\odot}$) at $z_{\mathrm{fg}} = \zfg$.  The predicted magnification maps from this analysis shows a large region of intermediate-to-high magnification ($\sim 6-15$ deg$^{2}$ with $\mu \geq 3$ for $z_{\mathrm{S}} = 10$).  However, without additional lensing constraints, several key parameters are unconstrained, including the halo ellipticity, orientation, and concentration.  An attempt to use the position of a multiply-imaged $z \approx 5.03$ source as a rudimentary constraint resulted in improved constraints on the location of the lensing critical curve near the lensed images, but a comprehensive treatment of the lensing constraints was beyond the scope of that work.

\subsection{Weak Lensing Analysis} \label{subsec:wl}
The effect of weak lensing on the background source population is characterized by the convergence, $\kappa$, and the shear, $\gamma$.  $\kappa$ represents the isotropic magnification due to lensing, and is defined as the projected surface mass density in units of the critical surface mass density for lensing, $\kappa \equiv \Sigma/\Sigma_{\mathrm{c}}$, where
\begin{equation} \label{eq:sigma_crit}
\Sigma_{\mathrm{c}} = \frac{c^{2}}{4 \pi \mathrm{G}} \frac{D_{\mathrm{S}}}{D_{\mathrm{L}} D_{\mathrm{LS}}}.
\end{equation}
$D_{\mathrm{L}}$, $D_{\mathrm{S}}$, and $D_{\mathrm{LS}}$ are the angular diameter distances to the lens, the source, and between the lens and source, respectively.  $\gamma$ represents a quadrupole anisotropy induced by lensing and can be observed from the ellipticities of the source population.  The complex shear $\gamma$ can be decomposed into the tangential component, $\gamma_{+}$, and the 45$^{\circ}$-rotated component, $\gamma_{\times}$.  In general, the observable quantity for weak lensing is not $\gamma$ but the reduced shear,
\begin{equation} \label{eq:reduced_shear}
g = \frac{\gamma}{1-\kappa}.
\end{equation}

\subsubsection{Selection of Background Source Galaxies} \label{subsubsec:wlselect}
We select background source galaxies for our weak lensing analysis using a $BRz^{\prime}$ color-color selection developed by \citet{medezinski2010}.  Selection of the background sources is key to accurate weak lensing measurements, as contamination by foreground sources or cluster members will dilute the lensing signal when unaccounted for, leading to a bias in the inferred cluster parameters \citep[e.g.,][]{medezinski2010,okabe2010}.

\begin{figure*}
\plottwo{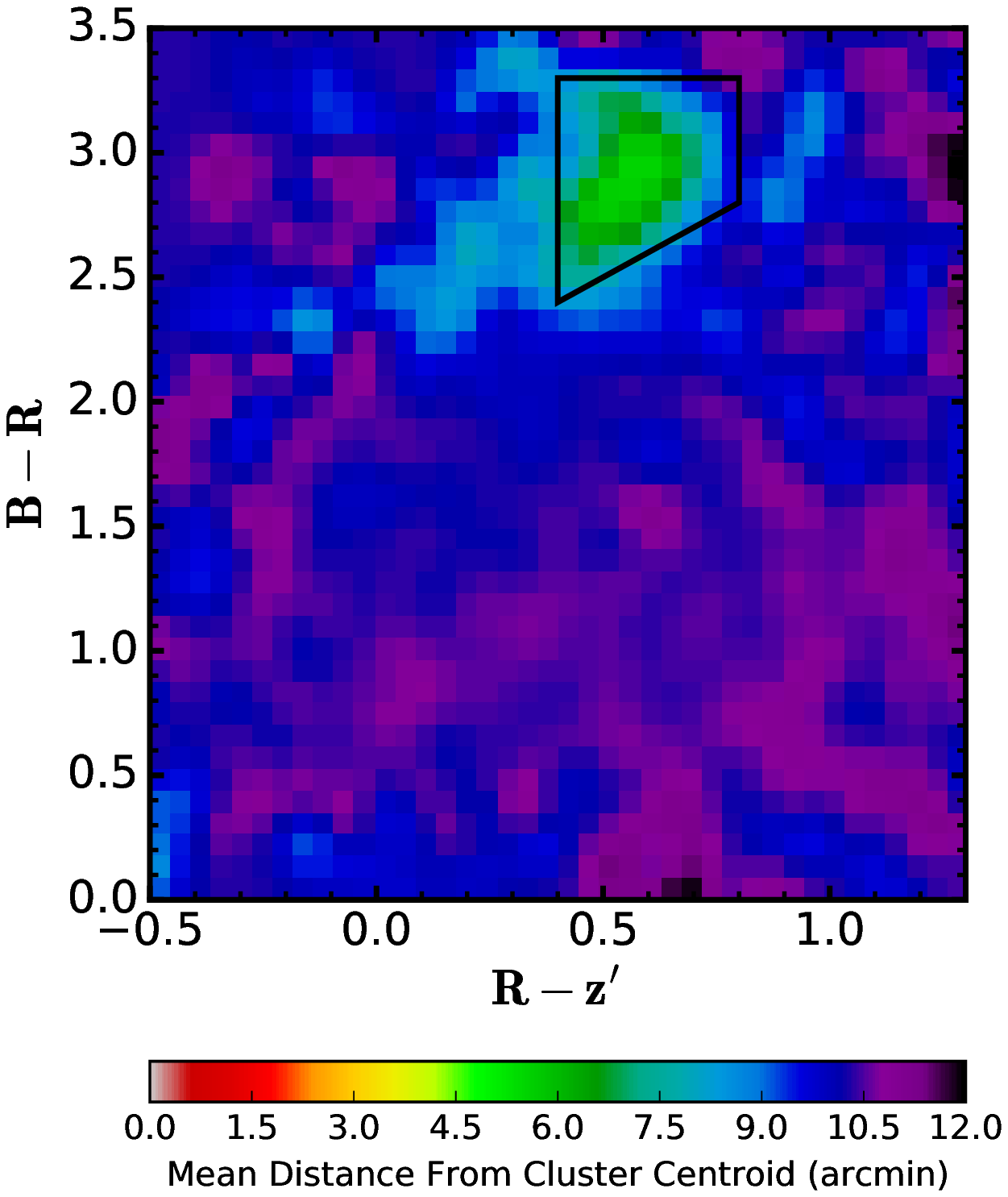}{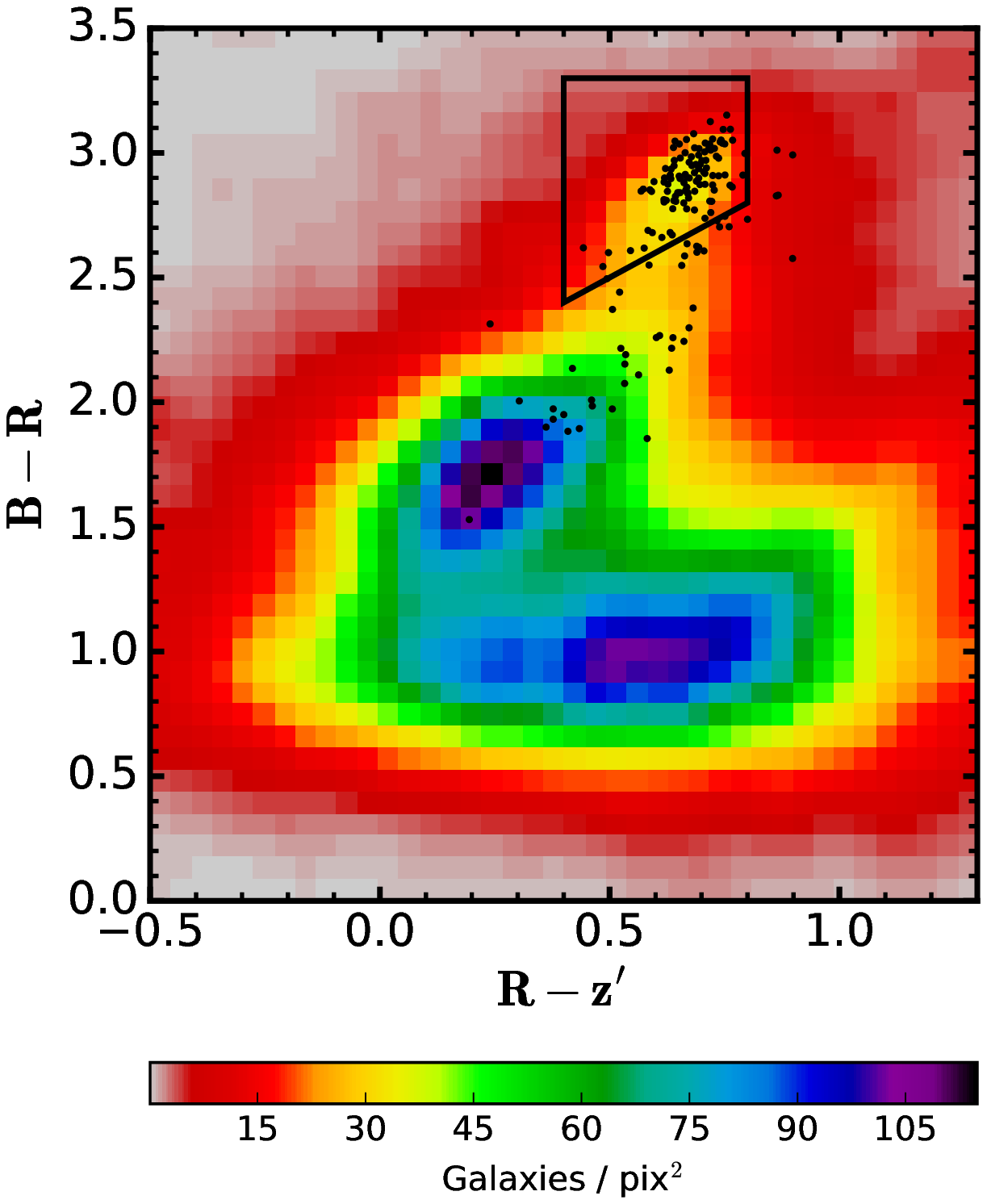}
\caption{{\bf Left:} $B-R$ vs. $R-z^{\prime}$ color for galaxies in the J0850 field.  The color of each cell indicates the mean angular offset from the cluster centroid for all galaxies in that particular location in color-color space.  The black polygon indicates our conservative selection of the region of color-color space corresponding to cluster member galaxies, as they will tend to be near the cluster centroid.
{\bf Right:} Number density in color-color space for galaxies in the field.  The black polygon is the same as in the left panel, and the enclosed galaxy concentration corresponds to likely cluster member galaxies, which we want to exclude from our source population.  The black points represent spectroscopically confirmed cluster members, the bulk of which are consistent with the selected overdensity.  Both figures have been smoothed with a 2D Gaussian filter for visualization purposes.
\label{fig:colordist}}
\end{figure*}

We start by placing all galaxies in the field on a $B-R$ vs. $R-z^{\prime}$ color-color diagram.  We divide this color-color space into discrete cells and calculate the mean projected distance of all galaxies from the cluster center \citep[as reported by][]{ammons2014} within each cell (Figure~\ref{fig:colordist}, left panel).  Figure~\ref{fig:colordist} shows a clear region where the mean projected distance of galaxies is smaller than the rest of the color-color space.  This region corresponds to colors of the cluster members themselves.  The number density of galaxies in the same color-color space (Figure~\ref{fig:colordist}, right panel) shows an overdensity in this same region, i.e., the massive cluster.  We want to exclude these galaxies from our source population.

\begin{figure*}
\plottwo{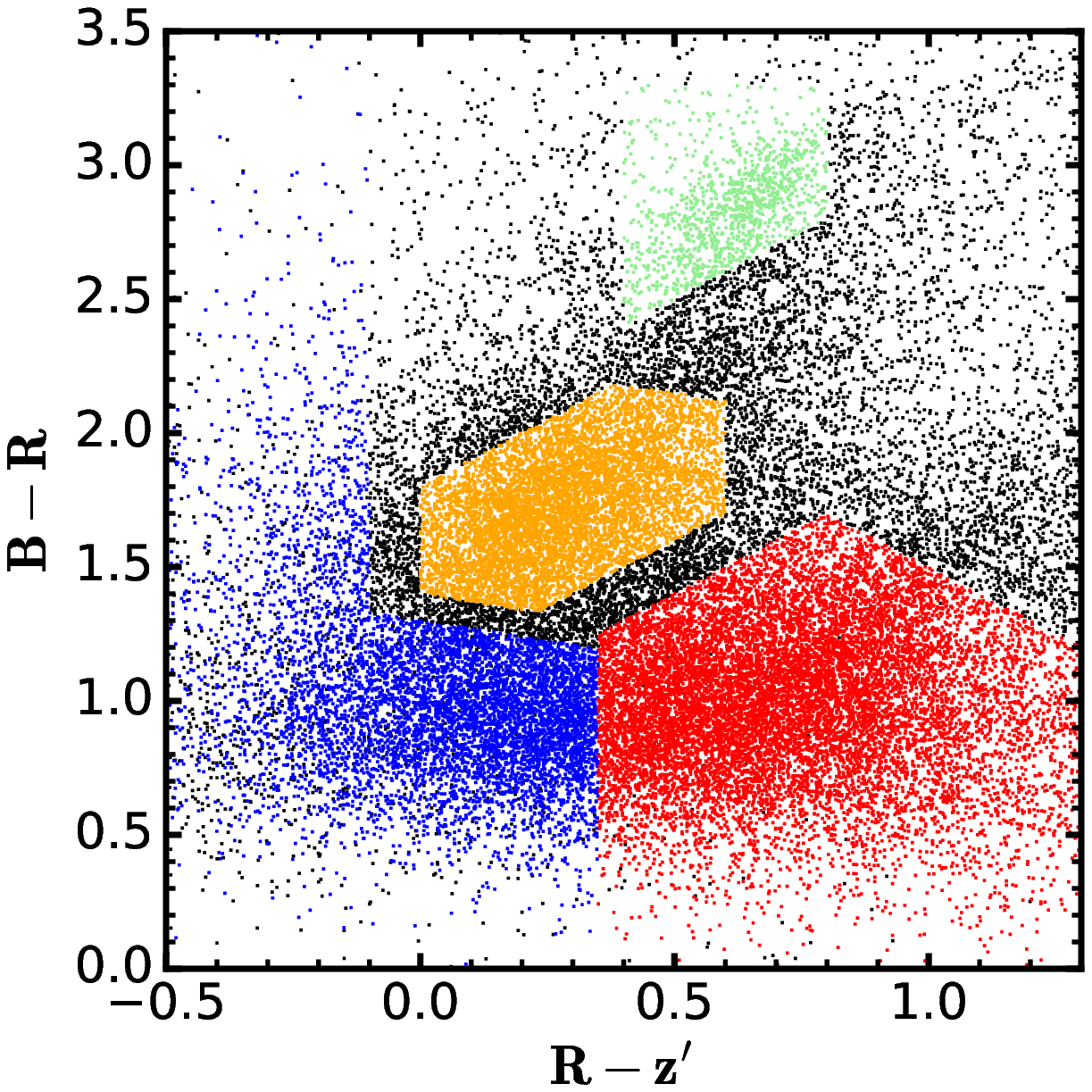}{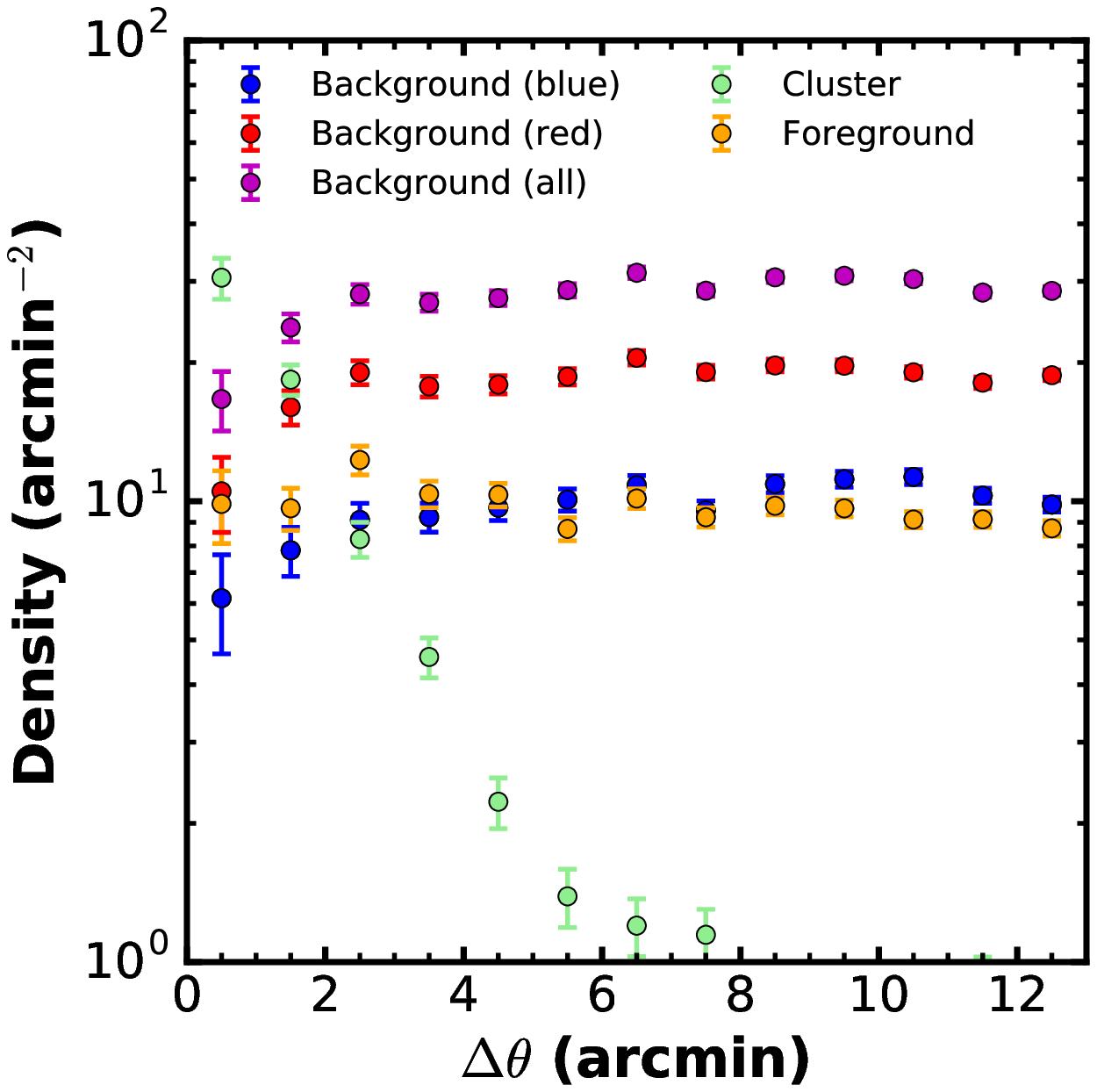}
\caption{{\bf Left:} $B-R$ vs. $R-z^{\prime}$ color for galaxies in the J0850 field.  The color of the points indicates galaxies photometrically selected to be cluster members (green), red background galaxies (red), blue background galaxies (blue), or foreground galaxies (orange).  Galaxies that are not selected by our cuts are shown in black.
{\bf Right:} Number density of galaxies as a function of distance from the cluster center.  The background densities have been corrected for masking effects due to cluster members, foreground objects, and defects.  The colors represent the same subsamples as in the left panel, except that black points indicate all background galaxies, both red and blue.  As expected, the number density of cluster galaxies increases sharply toward the center, while foreground galaxies maintain a roughly constant number density with distance.  We also see a decrease in the number density of background galaxies within the central $\sim2\arcmin$ as a result of depletion due to the lensing magnification.
\label{fig:colorselect}}
\end{figure*}

We define photometric selection criteria to select red background galaxies, blue background galaxies, cluster galaxies, and foreground galaxies.  Our criteria are listed in Table~\ref{tab:ccgals}.  The left panel of Figure~\ref{fig:colorselect} shows the individual galaxies in the field, color-coded by subsample (cluster, foreground, blue background, or red background).

\renewcommand*\arraystretch{1.5}
\begin{table*}
\caption{Color-Color Selected Galaxies \label{tab:ccgals}}
\begin{ruledtabular}
\begin{tabular}{c|cccc}
Sample &
Selection Criteria &
N &
$\langle z \rangle$\tablenotemark{a} &
$\langle \beta_{\mathrm{wl}} \rangle\tablenotemark{a} $
\\
\tableline
\multirow{4}{*}{Cluster} &
$z^{\prime} \leq 26.5$ &
\multirow{4}{*}{1147} &
\multirow{4}{*}{0.44} &
\multirow{4}{*}{\ldots}
\\
&
$B-R \leq 3.3$ &
&
&
\\
&
$0.4 \leq R-z^{\prime} \leq 0.8$ &
&
&
\\
&
$(B-R) - (R-z^{\prime}) \geq 2.0$ &
&
&
\\
\tableline
\multirow{2}{*}{Background (blue)} &
$22 \leq z^{\prime} \leq 26$ &
\multirow{2}{*}{7220} &
\multirow{2}{*}{1.53} &
\multirow{2}{*}{0.56}
\\
&
\{$R-z^{\prime} \leq 0.35$ AND $[(B-R) + 0.3\times(R-z^{\prime})] \leq 1.3$\} OR \{$R-z^{\prime} \leq -0.1$\}&
\\
\tableline
\multirow{4}{*}{Background (red)} &
$21 \leq z^{\prime} \leq 26$ &
\multirow{4}{*}{13593} &
\multirow{4}{*}{1.11} &
\multirow{4}{*}{0.56}
\\
&
$R-z^{\prime} \geq 0.35$ &
&
&
\\
&
$(B-R)-(R-z^{\prime}) \leq 0.9$ &
&
&
\\
&
$(B-R)+(R-z^{\prime}) \leq 2.5$ &
&
&
\\
\end{tabular}
\end{ruledtabular}
\tablenotetext{1}{Mean photometric redshift and distance ratio measured from COSMOS galaxies with same selection criteria applied.}
\end{table*}
\renewcommand*\arraystretch{1.0}

As a check of the robustness of our selection criteria, we plot the number density of the different galaxy subsamples as a function of distance from the cluster center (Figure~\ref{fig:colorselect}, right panel).  The cluster galaxy sample is concentrated near the cluster center and drops off with increasing radius, as expected.  The foreground galaxy subsample remains flat across the field, as those galaxies are not associated with the cluster and are not affected by lensing.  Both background source galaxy samples are flat at large radii, but show a decline within $\sim2\arcmin$ of the cluster center.  This is indicative of depletion of the galaxy number counts due to lensing magnification by the cluster \citep[e.g.,][]{broadhurst1995}.  When no source selection is applied, the cumulative galaxy number counts as a function of magnitude tend to show a power law behavior with a slope that is close to $s=0.4$, where no magnification bias due to lensing is expected.  However, owing to our color-color source selection, the count slope as a function of magnitude cut progressively decreases as we go to fainter magnitudes \citep[e.g.,][]{chiu2016}.  Because of the depth of the imaging data, the effective count slope at our limiting magnitude ($R_{c} = 27.5$) reaches $s \lesssim 0.2$, for which depletion of source counts is expected.

This depletion cannot be explained by the masking of background sources by cluster members, foreground objects, and defects (e.g., saturated stars and stellar trails), which is only a $\sim 10$\% effect in the central regions \citep[e.g.,][]{umetsu2016}. This masking effect is estimated and accounted for in the right panel of Figure~\ref{fig:colorselect}, following the procedure given in \citet[][see their Appendix A.2]{umetsu2011b} and \citet{umetsu2016}.

\begin{figure}
\plotone{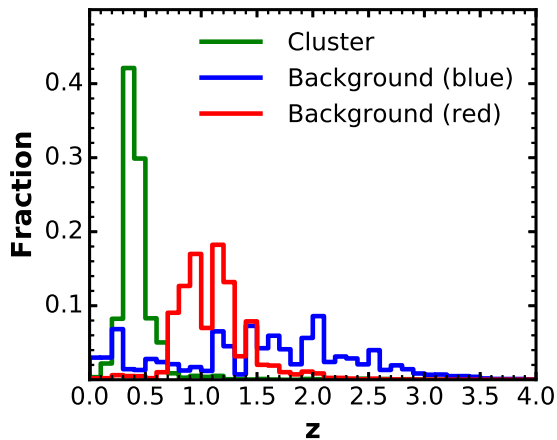}
\caption{Photometric redshift histogram of galaxies in the COSMOS catalog when applying the various selection criteria.  Shown are the redshift histograms for the cluster sample (green), blue background sample (blue), and red background sample (red).  The cluster subsample peaks around the cluster redshift, as expected, while the blue and red background subsamples mostly select galaxies at higher redshifts.
\label{fig:zhist}}
\end{figure}

The weak lensing signal scales with the angular diameter distance ratio $\beta_{\mathrm{wl}} \equiv D_{\mathrm{LS}}/D_{\mathrm{S}}$, with $\beta_{\mathrm{wl}} \equiv 0$ for foreground objects with $z < z_{\mathrm{L}}$.  For statistical weak lensing measurements, the mean distance ratio is
\begin{equation} \label{eq:beta_wl}
\langle \beta_{\mathrm{wl}} \rangle = \frac{\int_{0}^{\infty} N(z) \beta_{\mathrm{wl}}(z) dz}{\int_{0}^{\infty} N(z) dz},
\end{equation}
where $N(z)$ is the redshift distribution function of the background population.  We estimate and correct the depths of the different subsamples by applying similar selection criteria to the Cosmic Evolution Survey \citep[COSMOS;][]{capak2007}, for which accurate photometric redshifts have been derived from 30-band photometry \citep{ilbert2009}.  Since the COSMOS filters do not include the $R_{c}$ band, we calculate fluxes from the best-fit templates with EAZY \citep{brammer2008}.  We then calculate the mean redshifts and distance ratios for the background subsamples (Table~\ref{tab:ccgals}).  The redshift histograms of the various galaxy subsamples are shown in Figure~\ref{fig:zhist}.  The mean redshifts of the red and blue background subsamples are $\langle z_{\mathrm{red}} \rangle = 1.11$ and $\langle z_{\mathrm{blue}} \rangle = 1.53$, respectively.  Despite this difference, these two subsamples coincidentally have the same mean distance ratio of $\langle \beta_{\mathrm{wl}} \rangle = 0.56$ due to the fact that the blue population has a larger fraction of low redshift ($\beta_{\mathrm{wl}} = 0$) interlopers.  We use the value $\langle \beta_{\mathrm{wl}} \rangle = 0.56 \pm 0.03$ for our weak lensing analysis, which accounts for a typical 5\% uncertainty \citep[e.g.,][]{umetsu2014} that is marginalized over in the modeling.  Updated photometric redshifts from \citet{laigle2016} are available in the COSMOS field, but using them only changes the mean distance ratio to $\langle \beta_{\mathrm{wl}} \rangle = 0.60$, which is still consistent with our previous estimate.

We exclude background galaxies in a 1\arcmin~radius region around the cluster centroid determined by \citet{ammons2014} when performing the weak lensing analysis, as sources there may be in either the strong lensing or highly non-linear regime.  This central region also has a higher density of cluster member galaxies, where residual contamination is more likely.

\subsubsection{Weak Lensing Methodology} \label{subsubsec:wlmethod}
We use a weak-lensing analysis pipeline based on a modified version the IMCAT software package \citep{kaiser1995} to perform shape measurements of the background galaxies. Full details of the formalism, including the shear calibration method, are described in \citet{umetsu2010,umetsu2012,umetsu2014}, and its implementation has been applied extensively to cluster weak-lensing studies with Subaru/Suprime-Cam observations \citep[e.g.,][] {medezinski2010,medezinski2011,medezinski2013,medezinski2016,umetsu2008,umetsu2009,umetsu2010,umetsu2011a,umetsu2011b,umetsu2012,umetsu2014,umetsu2015,zitrin2011b,coe2012,wegner2017}. In this work, we follow the analysis procedures that \citet{umetsu2014} employed for the CLASH survey \citep{postman2012}.  Briefly summarizing, the procedures include \citep[see also Section 3 of][]{umetsu2016}: (1) object detection using the IMCAT peak finder, {\sc hfindpeaks}, (2) conservative close-pair rejection to reduce the crowding and deblending effects, and (3) shear calibration developed by \citet{umetsu2010} to minimize the inherent noise bias.

Using simulated Subaru/Suprime-Cam images \citep{massey2007,oguri2012}, \citet{umetsu2010} find that the shear signal can be
recovered with $|m|\sim 0.05$ of the multiplicative calibration bias
and $c\sim 10^{-3}$ of the residual shear offset \citep[as defined by][]{heymans2006,massey2007}. Accordingly, we include for
each galaxy a shear calibration factor of $g\to g/0.95$ (see Equation~\ref{eq:reduced_shear}) to account for residual calibration. As discussed by \citet{umetsu2012}, $m$ depends modestly on the width and quality of the
point-spread function (PSF). This variation with the PSF properties
limits the accuracy of shear calibration to $\delta m\sim 0.05$ \citep{umetsu2012}.

We use the $R_{c}$ band for the weak lensing analysis, as it has the best combination of depth and seeing conditions.  We exclude frames where the seeing is $> 0.9\arcsec$ and do not match the PSF across frames before coadding to the final stacked frame.  The median seeing of the final data frame is 0\farcs81.

The mass centroid of the cluster is allowed to vary and is given a Gaussian prior centered on the position of the main cluster halo determined by \citet{ammons2014} based on the mean position of the spectroscopically confirmed cluster members.  The 1$\sigma$ width of the prior is 12\arcsec, which is the uncertainty on the \citet{ammons2014} centroid position estimated from a bootstrap resampling of the cluster galaxies.  The halo is assumed to be a Navarro-Frenk-White \citep[NFW;][]{navarro1996} profile with log-uniform priors on its mass and concentration, which are appropriate for positive-definite quantities \citep[e.g.,][]{feroz2008,sereno2013,umetsu2014}.  The halo ellipticity $\epsilon$ (defined as $\epsilon \equiv 1-b/a$, where $b/a$ is the minor-to-major projected axis ratio) and orientation $\theta_{\epsilon}$ are given uniform priors.  We account for the foreground halo at $z = \zfg$ by fixing its centroid to the coordinates determined from \citet{ammons2014} and assuming that it is a spherical NFW profile.  Its mass is allowed to vary and is given a log-uniform prior.  The foreground halo's concentration is assumed to vary monotonically with mass according to the mass-concentration relation of \citet{dutton2014}.

\subsection{Strong Lensing Analysis} \label{subsec:sl}
\begin{figure}
\plotone{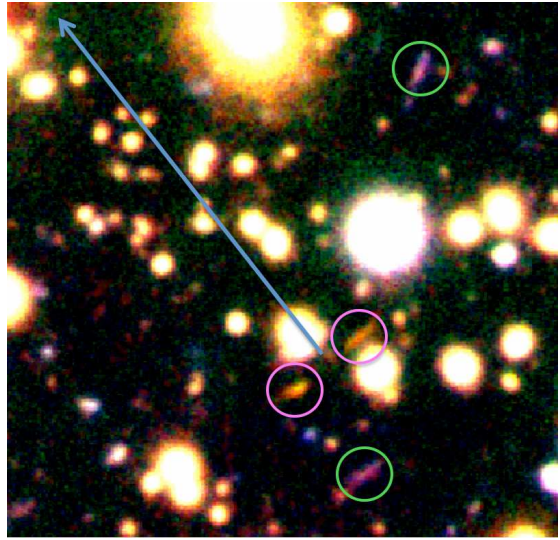}
\caption{Subaru/Suprime-Cam multicolor image of a $1\arcmin \times 1\arcmin$ portion of the J0850 field.  The magenta circles show two images of a strongly-lensed background source that has a photometric redshift of $z = 5.03$.  The blue arrow points to the cluster centroid as determined from the spectroscopically confirmed members. The morphologies, direction of elongation, and distance
from the cluster centroid ($\sim50\arcsec$) suggest that they are highly magnified.  The green circles show other candidates lensed arcs in the field.  Figure reproduced from \citet{ammons2014}.
\label{fig:arcs}}
\end{figure}

We perform our strong lensing analysis using a parametric source reconstruction method with a multiply-imaged background galaxy as the constraint. This galaxy was revealed through multiband Suprime-Cam imaging and has a photometric redshift of $z = 5.03_{-0.17}^{+0.21}$ \citep{ammons2014}.  The source is lensed into two small arcs separated by $\sim8\arcsec$ that are located roughly $\sim50\arcsec$ to the southwest of the cluster center (Figure~\ref{fig:arcs}).  These arcs, given their small separation, strongly constrain the critical curve in this region.

\subsubsection{Construction of Mass Models} \label{subsubsec:slmassmod}
Using the posterior distribution of the cluster properties from our weak lensing analysis, we generate 100,000 Monte Carlo realizations of the mass distribution in the field, including the main cluster halo, the foreground halo, and the cluster and LOS galaxies.  The main cluster is given a random triaxiality and 3D orientation such that its projected ellipticity and orientation matches that of the weak lensing results for a particular model.  Our analysis accounts for line-of-sight effects using the full multi-plane lens equation \citep[e.g.,][]{blandford1986,kovner1987,schneider1992,petters2001,collett2014,mccully2014}, as ignoring LOS structure can lead to biases in the inferred model parameters and resulting magnification maps \citep[e.g.,][]{bayliss2014}.

The procedure for constructing the full LOS mass distribution in this manner is described in \citet{ammons2014}, with some small modifications detailed here.  The virial mass of the main cluster is divided among a common NFW dark matter halo and the spectroscopically confirmed member galaxies, which are assumed to be truncated singular isothermal spheres \citep[see][]{wong2011}.  The foreground group's mass is also apportioned between a common dark matter halo and confirmed member galaxies in a similar way.  In addition to the spectroscopically observed galaxies, we add galaxies with photometric redshifts from SDSS to our mass model by selecting those galaxies brighter than $i < 21.1$ and within $1\farcm5$ of the field center.  Galaxies with $|z - z_{\mathrm{L}}|/(1+z_{\mathrm{L}}) \leq 0.1$ are treated as  members of the main cluster, and their redshifts are fixed to the cluster redshift.  For the remaining galaxies, we assign redshifts from a Gaussian distribution centered on their SDSS photometric redshift and with a 1$\sigma$ width equal to their photo-z uncertainty.  Redshifts are drawn from these distributions for each Monte Carlo realization so that photo-z errors are accounted for in our modeling.  We also include two galaxies in close proximity to the lensed arcs that are further from the cluster center and not in the spectroscopic catalog.  These galaxies appear blended in the Suprime-Cam imaging data.  We use {\sc GALFIT} \citep{peng2002} to deblend them to determine their relative fluxes, then scale their magnitudes proportionally such that the sum of their fluxes matches the $i$-band photometry from SDSS.  The two galaxies are treated as cluster members, as the color inferred from the blended photometry is consistent with that of the spectroscopically-confirmed cluster galaxies.

\subsubsection{Application of Strong Lensing Constraints} \label{subsubsec:slapp}
\begin{figure*}
\plotone{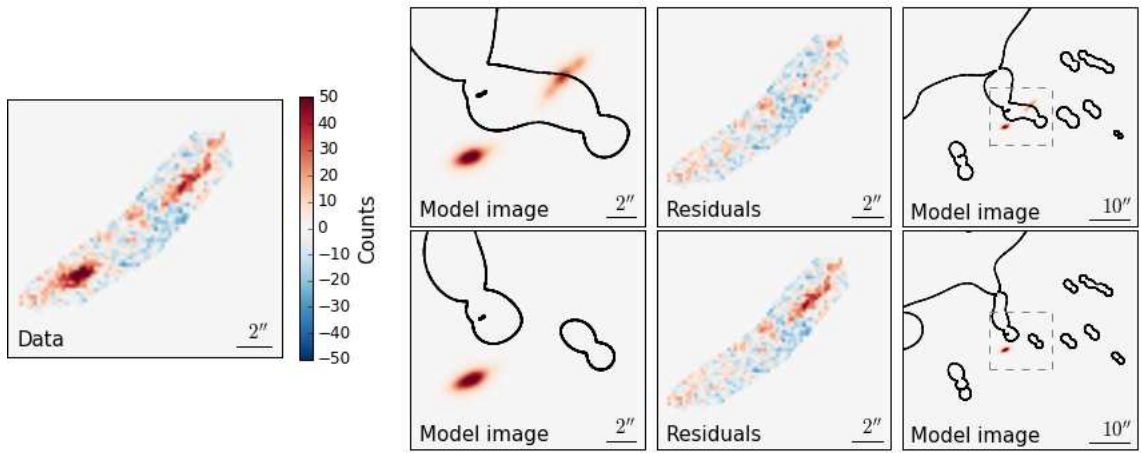}
\caption{
{\bf Left:} Data containing the two arclets used in the strong lensing analysis. Note that a mask has been applied so that galaxies nearby are not included in the reconstruction.
{\bf Second column:}  Images (red) and critical curves (black) predicted by lens models.  The images have been blurred by the PSF.
{\bf Third column:} Residuals between the observed and model images.
{\bf Fourth column:}  Larger map to show the predicted images in relation to the cluster-scale critical curve; the grey dashed box indicates the region shown in the second column.
The top row shows results for a successful model that reproduces both arcs, while the bottom row shows a model that fails to reproduce two arcs.
\label{fig:slresids}}
\end{figure*}

We use the two images of the lensed background galaxy as constraints for a strong lensing analysis.  The source is assumed to be at a fixed redshift of $z=5.03$.  The uncertainty in the photometric redshift estimate is unlikely to matter because the lensing properties of a cluster at low and intermediate redshifts are fairly insensitive to source redshift beyond $z_{\mathrm{S}}\sim4$. We use the {\sc pixsrc} software \citep{tagore2014}, an extension to {\sc lensmodel} \citep{keeton2001} that performs a pixelated reconstruction of the arcs using the full surface brightness distribution in the $R_{c}$ band.  Due to a lack of discernible structure within the arcs, we assume the source follows a S\'{e}rsic profile with $n=1$, although our results are robust to different choices of $n$. The position, brightness, scale radius, ellipticity, and orientation of the source are optimized so that the lensed images, convolved with a PSF that mimics seeing conditions during observations, most closely match the data (as measured by the $\chi^2$ sum over residuals).  The strong lensing analysis is depicted in Figure~\ref{fig:slresids}.

\begin{figure}
\plotone{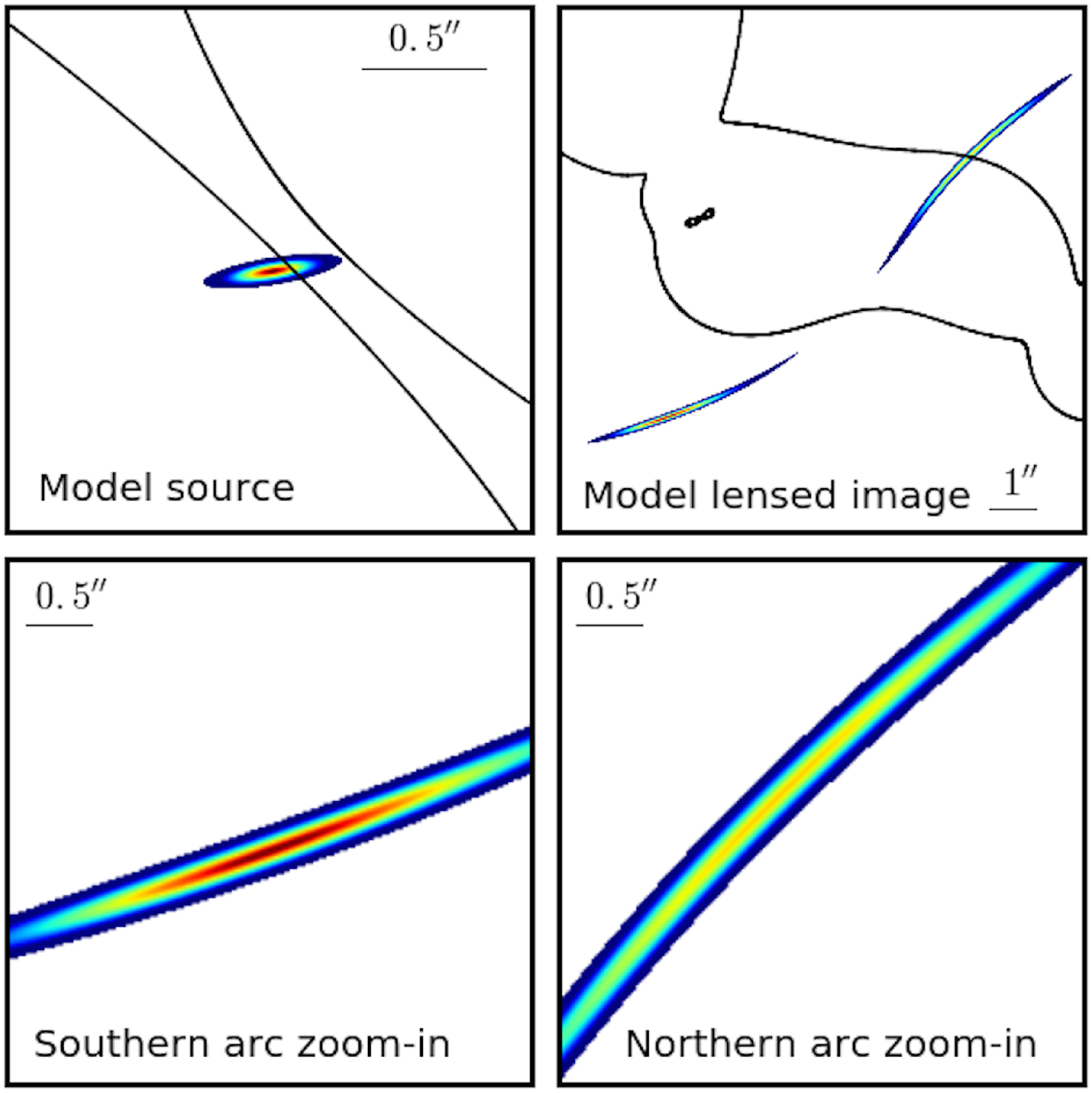}
\caption{
{\bf Top left:} Source and caustics for the best strong lensing model. The core of the source galaxy lies just outside of the caustic, so it is not multiply-imaged, but the outskirts of the galaxy fall within the caustic and produce the northern arc.
{\bf Top right:} Lensed images and critical curves for the same model. Here we have not convolved with the PSF in order to see the intrinsic structure of the images. Since lensing conserves surface brightness, intensities in the source plane map directly to those in the image plane.
{\bf Bottom:} Zoom-in on the individual arcs. The core of the source galaxy is seen only in the brighter southern arc (bottom left panel), whereas the fainter northern arc (bottom right panel) features two images of the outskirts of the source galaxy that merge together at the critical curve.
\label{fig:slmodel}}
\end{figure}

Using the full pixel information differs from standard modeling methods which use only the centroid positions of images as constraints.  This technique is advantageous in this case where the lensed images are near critical curves (implying that the source is close to a caustic), and information from the extended images can be used.  Figure~\ref{fig:slmodel} shows our best fit model in detail.  This model predicts the core of the source galaxy to lie outside the caustic, so that it is seen only in the southern arc and does not produce multiple images.  A second arc is visible because part of the source overlaps the caustic; that part is stretched into two additional images that merge together to create the northern arc.  In other words, most parts of the source yield just one image, but the western outskirts actually create three images where two of them meet at the critical curve to become the northern arc.  A standard modeling approach that treats the arcs as two point-like images would not capture the full complexity of the lensing.

\begin{figure}
\plotone{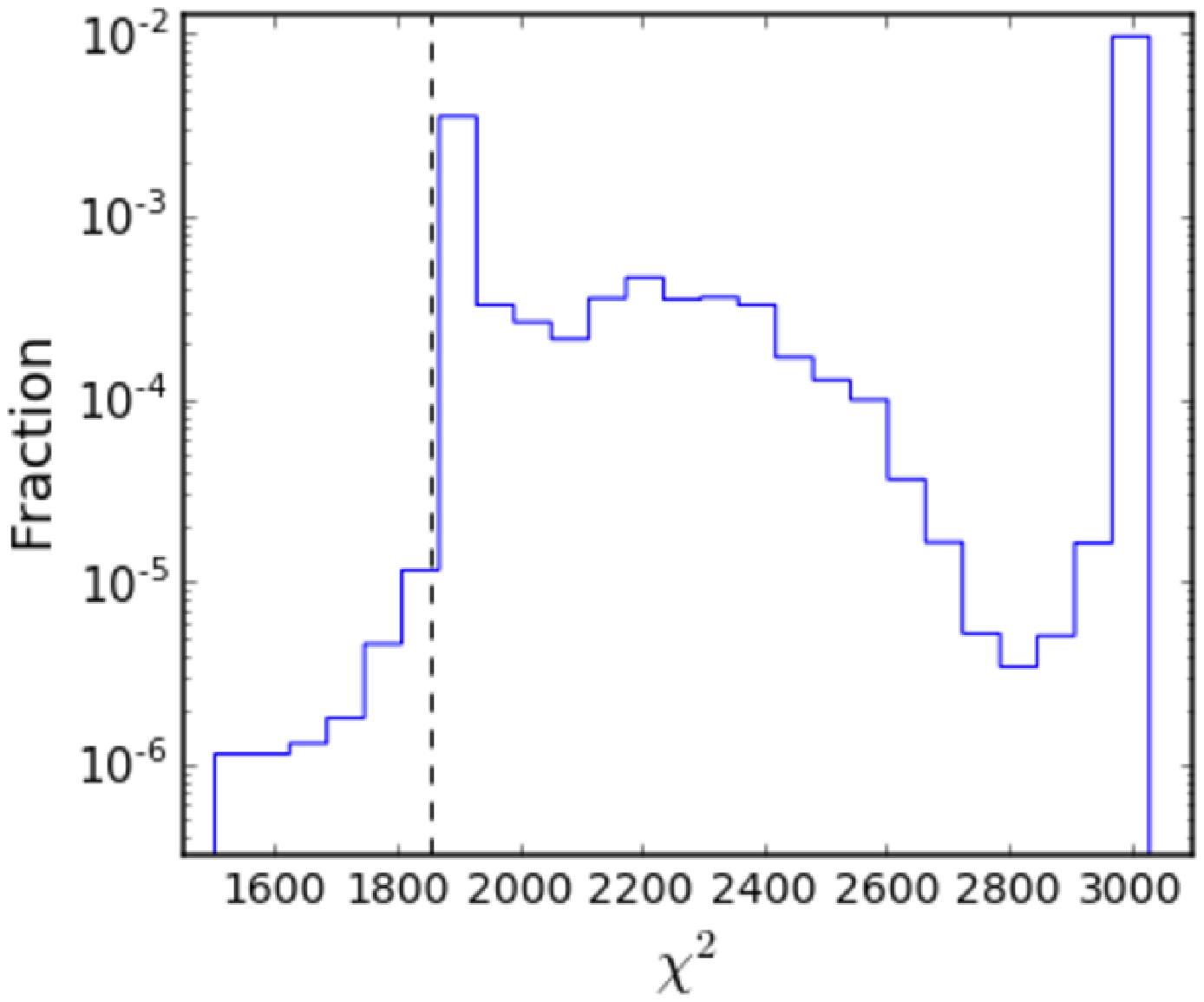}
\caption{A histogram of $\chi^2$ values measured in the strong lensing analysis; this is the full (not reduced) $\chi^2$ (see text for comments about interpreting $\chi^2$ values).  The large jump at $\chi^2 \simeq 1855$ indicates a transition from models that can produce two arcs (e.g., the top row in Figure~\ref{fig:slresids}) into models that cannot reproduce the fainter northern arc (e.g., the bottom row in Figure~\ref{fig:slresids}).  We therefore impose a cut indicated by the dashed line.  The spike at $\chi^2 \simeq 3000$ is due to models that fail to reproduce either of the arcs.
\label{fig:slchi2}}
\end{figure}

To compute $\chi^2$ values from image residuals, we measure a noise level of $\sigma_{\mathrm{noise}} = 9$ counts in a galaxy-subtracted image near but not overlapping the observed arcs.  Applying this noise level to all 100,000 Monte Carlo realizations of the mass distribution yields the $\chi^2$ histogram in Figure~\ref{fig:slchi2}.  The best model has $\chi^2 = 1502$ for $\mathrm{N_{pix}} = 1011$ pixels.  The residuals do not have obvious structure (see the top row of Figure~\ref{fig:slresids}), and the pixel distribution is roughly Gaussian with a mean of $-2.0$ and a standard deviation of $10.8$.  We conjecture that the noise properties vary slightly between the region where we measure the noise and the location of the arcs due to statistical variations and/or imperfect galaxy subtraction.  We keep this in mind when interpreting $\chi^2$ values but do not attempt to rescale the noise.

The vast majority of models cannot reproduce two arcs because the critical curves are not in the right place as seen in the bottom row of Figure~\ref{fig:slresids} and in Figure~\ref{fig:slchi2} (the large jump at $\chi^2 \approx 1855$).  Only 107 of the models lie below this threshold and are able to reproduce the two arcs.  We consider these models successful according to the strong lensing analysis.  When deriving parameter constraints (Section~\ref{subsec:slresults}), we treat all of the successful strong lensing models with equal weight.  We do not weight by likelihood ${\cal L} \propto e^{-\chi^2/2}$ due to questions about the noise level and because our assumption of a single S\'{e}rsic source with fixed index is probably too simplistic (limited by current data).  We emphasize that these issues do not affect whether models can produce two arcs, nor does changing the $\chi^2$ threshold significantly affect the derived parameter constraints.

\section{Results} \label{sec:results}

\subsection{Constraints from Weak Lensing Analysis} \label{subsec:wlresults}
\begin{figure}
\plotone{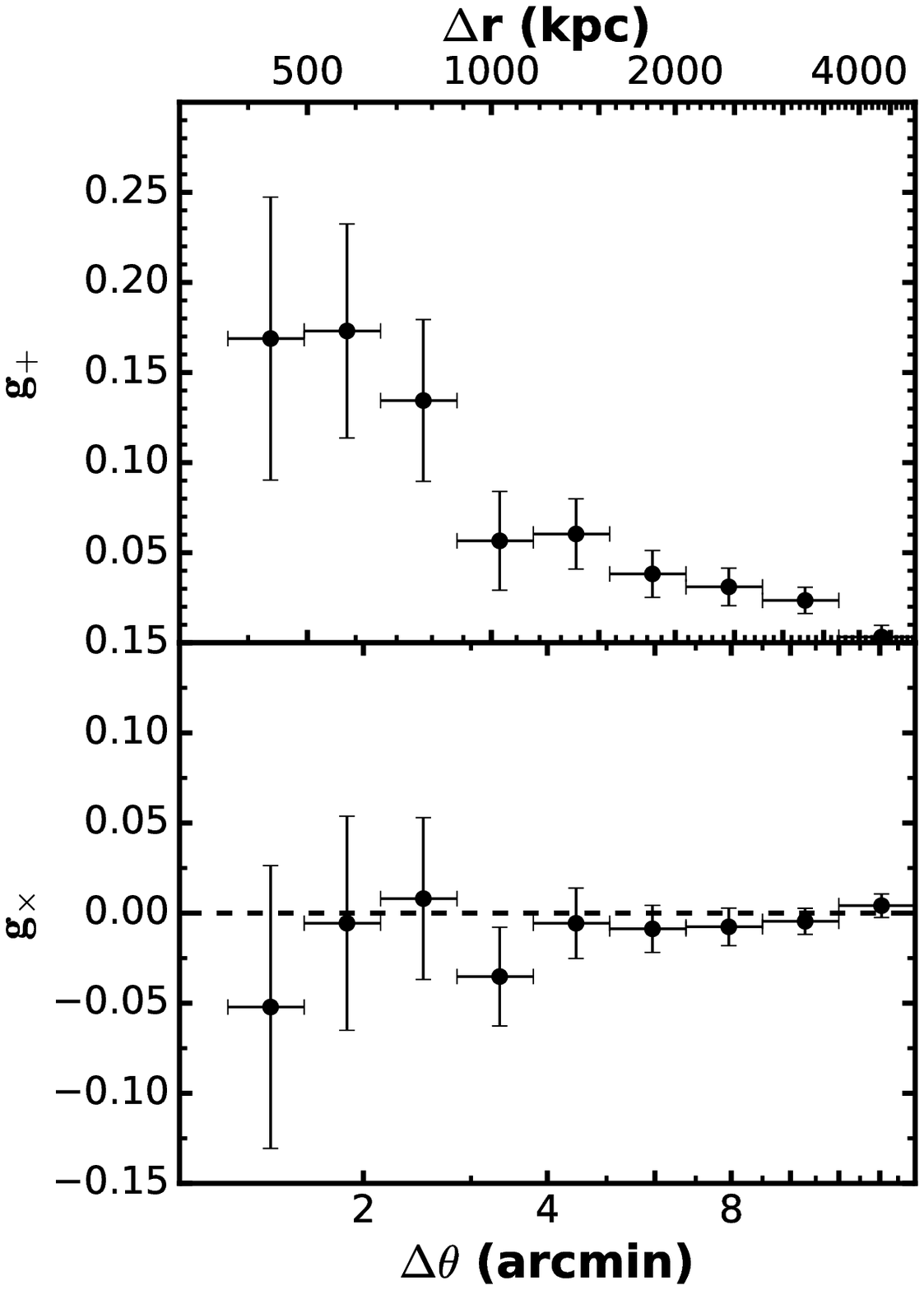}
\caption{{\bf Top:} Azimuthally averaged tangential reduced shear ($g_{+}$) profile as a function of distance from the cluster center.  The rising trend of $g_{+}$ at smaller radius demonstrates the robustness of our selection of background galaxies.
{\bf Bottom:} Azimuthally averaged $45^{\circ}$-rotated reduced shear ($g_{\times}$) profile as a function of distance from the cluster center.  The $g_{\times}$ component is consistent with a null detection in all bins, as is expected in the absence of systematic errors in the shape measurements.
\label{fig:wlshear}}
\end{figure}

In Figure~\ref{fig:wlshear}, we plot the azimuthally averaged profile of the tangential reduced shear ($g_{+}$) and $45^{\circ}$-rotated reduced shear ($g_{\times}$) as a function of distance from the cluster center reported by \citet{ammons2014}.  In the weak lensing regime, the shear field should be curl-free, so the presence of the $g_{\times}$ component can be used as a check for systematic errors in the shape measurements.  Here, the $g_{+}$ component rises toward the cluster center, as expected, and the $g_{\times}$ component is consistent with zero for all bins.

\begin{figure}
\plotone{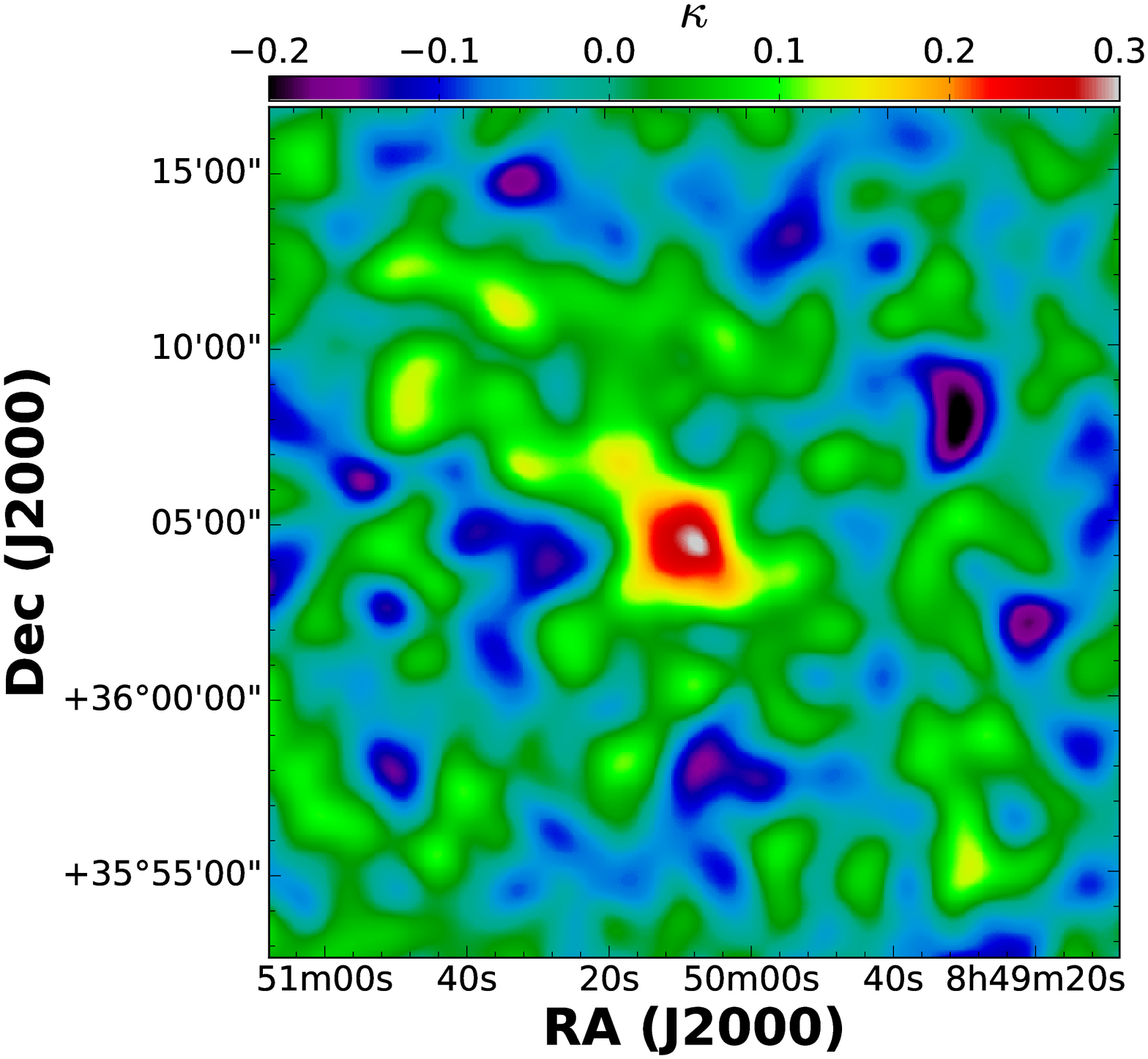}
\caption{Convergence ($\kappa$) map from our weak lensing analysis.  For visualization purposes, the map is smoothed with a circular Gaussian of FWHM 1\farcm5. The field of view is $24\arcmin \times 24\arcmin$ (corresponding to $\sim7.4\times7.4$ Mpc at the lens redshift) and is centered on the cluster.  The color bar indicates the $\kappa$ at each point in the field.  The massive cluster is detected as the peak near the field center.  The smaller foreground group is undetected, as its mass is below the scale that can be probed by the weak lensing.\label{fig:kappamap}}
\end{figure}

Figure~\ref{fig:kappamap} shows the convergence ($\kappa$) map derived from our weak lensing analysis using the \citet{kaiser1993} linear inversion method \citep[see][]{umetsu2009}.  The massive cluster is clearly detected near the field center.  The much smaller foreground halo is undetected, as its mass is below the scale that can be probed by the weak lensing.

\begin{figure*}
\plotone{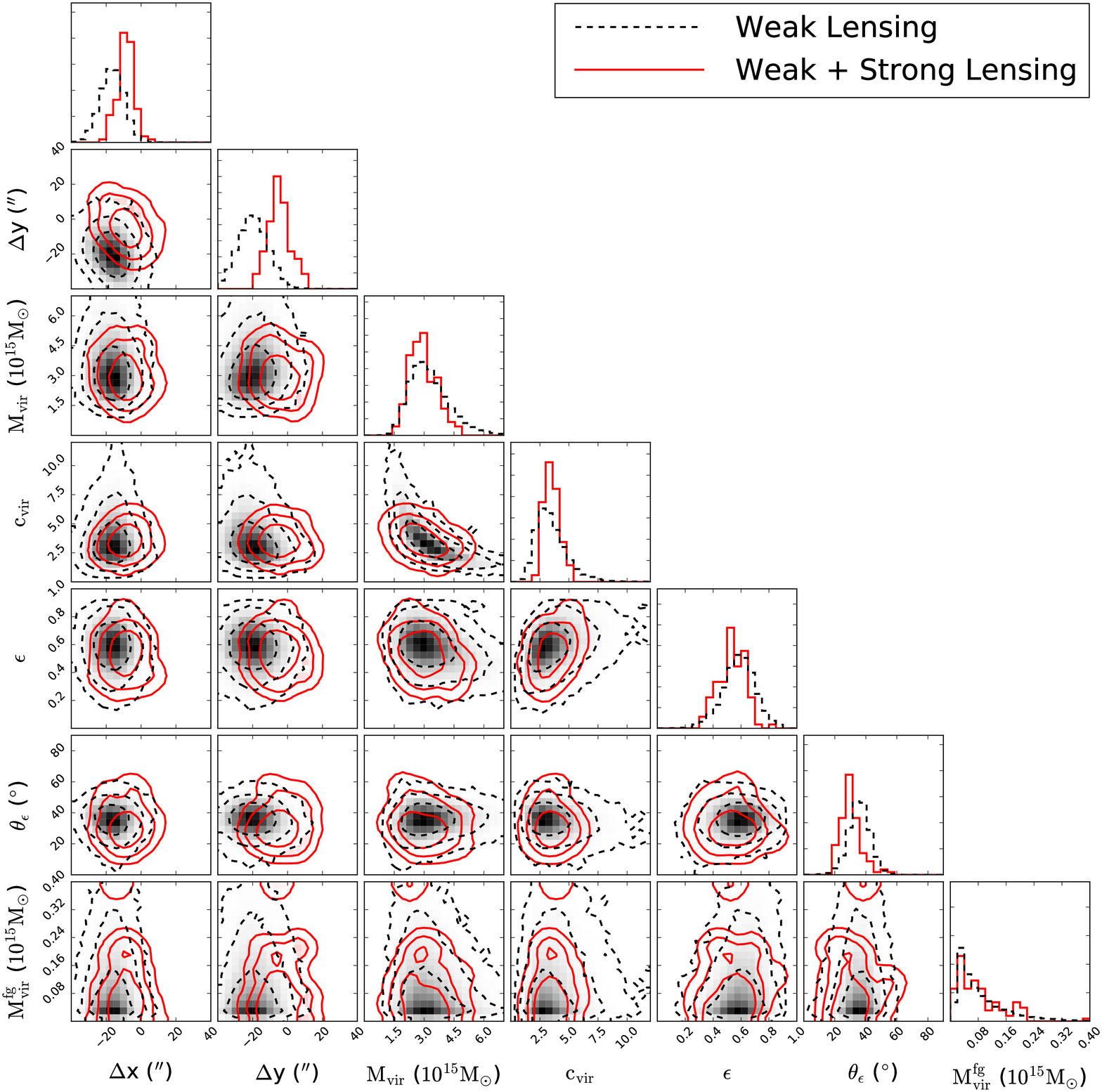}
\caption{Posterior parameter distributions and covariances for the main cluster halo from our lensing analysis.  The contours represent 68\%, 95\%, and 99.7\% quantiles.  The plots along the diagonal show marginalized distributions of the various parameters.  The dotted black contours and histograms are the weak lensing constraints alone, while the red solid contours and histograms are the combined weak and strong lensing constraints.  Shown (from left to right, top to bottom) are the $\Delta$x and $\Delta$y positions of the main cluster centroid relative to that determined by \citet{ammons2014}, the main cluster's virial mass, concentration, ellipticity, position angle (measured East of North), and the foreground halo's virial mass.  The addition of the strong lensing information tightens the constraints on the main cluster's mass, concentration, and centroid.
\label{fig:modelparams}}
\end{figure*}

We show the posterior distributions of parameters from our weak lensing analysis in Figure~\ref{fig:modelparams}.  The priors and posterior constraints are also given in Table~\ref{tab:modelparams} (center column).  The virial quantities are defined using the overdensity criterion of \citet{bryan1998} for our assumed cosmology.

The main cluster at $z = \zl$ has a virial mass of $\mathrm{M_{vir}} = \mclwl$, which is in good agreement with that determined independently from dynamics by \citet{ammons2014}.  The halo concentration is consistent with that of other clusters of comparable mass and redshift \citep[e.g.,][]{umetsu2016}, but has a greater ellipticity \citep[e.g.,][]{despali2017}, with the major axis roughly pointing toward the multiply-imaged arcs.  This high ellipticity ($\epsilon = 0.59_{-0.12}^{+0.11}$) suggests that the main J0850 cluster is an efficient lens (i.e., there should be a high number density of multiple images) in comparison to a spherical halo with an equivalent area enclosed within the critical curves \citep{zitrin2013}.

The foreground halo at $z = \zfg$ has a virial mass of $\mathrm{M_{vir}^{fg}} = \mfgwl$.  The large uncertainties arise because its influence is below the scale probed by the weak lensing, as was seen in Figure~\ref{fig:kappamap}.  However, the weak lensing analysis does place a rough upper limit on its mass, as higher mass halos would influence the external shear field and thus be detected.

\renewcommand*\arraystretch{1.5}
\begin{table*}
\caption{J0850 Lensing Parameter Constraints \label{tab:modelparams}}
\begin{ruledtabular}
\begin{tabular}{l|ccc}
Parameter &
Prior &
Weak Lensing Posterior &
Weak+Strong Lensing Posterior
\\
\tableline
$\Delta$x (\arcsec)\tablenotemark{a} &
Gaussian; $x_{0} = 0.0$; $\sigma=12.0$ &
$-16.4_{-6.8}^{+6.5}$ &
$-8.8_{-4.1}^{+4.1}$
\\
$\Delta$y (\arcsec)\tablenotemark{a} &
Gaussian; $y_{0} = 0.0$; $\sigma=12.0$ &
$-20.0_{-8.4}^{+8.9}$ &
$-4.7_{-6.4}^{+5.4}$
\\
$\mathrm{M_{vir}}$ ($10^{15}~\mathrm{M_{\odot}}$) &
Log-uniform; [0.01, 10] $h_{100}^{-1}$ &
$3.15_{-0.81}^{+1.13}$ &
$2.93_{-0.65}^{+0.71}$
\\
$\mathrm{c_{vir}}$ &
Log-uniform; [0.01, 20] &
$3.26_{-1.01}^{+1.47}$ &
$3.46_{-0.59}^{+0.70}$
\\
$\epsilon$ &
Uniform; [$0, 0.9$] &
$0.59_{-0.12}^{+0.11}$ &
$0.53_{-0.10}^{+0.09}$
\\
$\theta_{\epsilon}$ ($^{\circ}$)\tablenotemark{b} &
Uniform; [$-90$, $90$] &
$35.7_{-6.6}^{+6.6}$ &
$29.5_{-5.6}^{+6.3}$
\\
\tableline
$\mathrm{M_{vir}^{fg}}$ ($10^{15}~\mathrm{M_{\odot}}$) &
Log-uniform; [0.01, 10] $h_{100}^{-1}$ &
$6.22_{-3.62}^{+9.21} \times 10^{-2}$ &
$6.53_{-4.02}^{+9.24} \times 10^{-2}$
\\
\end{tabular}
\end{ruledtabular}
\tablecomments{Reported values are medians, with errors corresponding to the 16th and 84th percentiles.}
\tablenotetext{1}{Offset of main cluster centroid relative to that determined by \citet{ammons2014}.}\tablenotetext{2}{Position angle is measured East of North.}\end{table*}
\renewcommand*\arraystretch{1.0}

\subsection{Combined Weak and Strong Lensing Constraints} \label{subsec:slresults}
We show the constraints from the combined strong and weak lensing analysis in Table~\ref{tab:modelparams} (right column).  The distributions are also plotted in Figure~\ref{fig:modelparams} for comparison with the constraints from weak lensing alone.  The combined analysis produces a main cluster virial mass of $\mathrm{M_{vir}} = \mclsl$, with a tighter uncertainty than from the weak lensing constraints alone.  The inferred virial mass is consistent with the dynamical mass estimate of \citet{ammons2014}, confirming that this is among the most massive known lensing clusters.  The mass of the foreground halo, which has an upper limit from the weak lensing analysis, unsurprisingly does not change much with the addition of the strong lensing constraint from a multiple image pair formed by the main cluster.  The galaxy redshift distribution in the J0850 field may contain smaller structures along the line of sight \citep{ammons2014}, but they are similarly insensitive to the lensing constraints.

\begin{figure}
\plotone{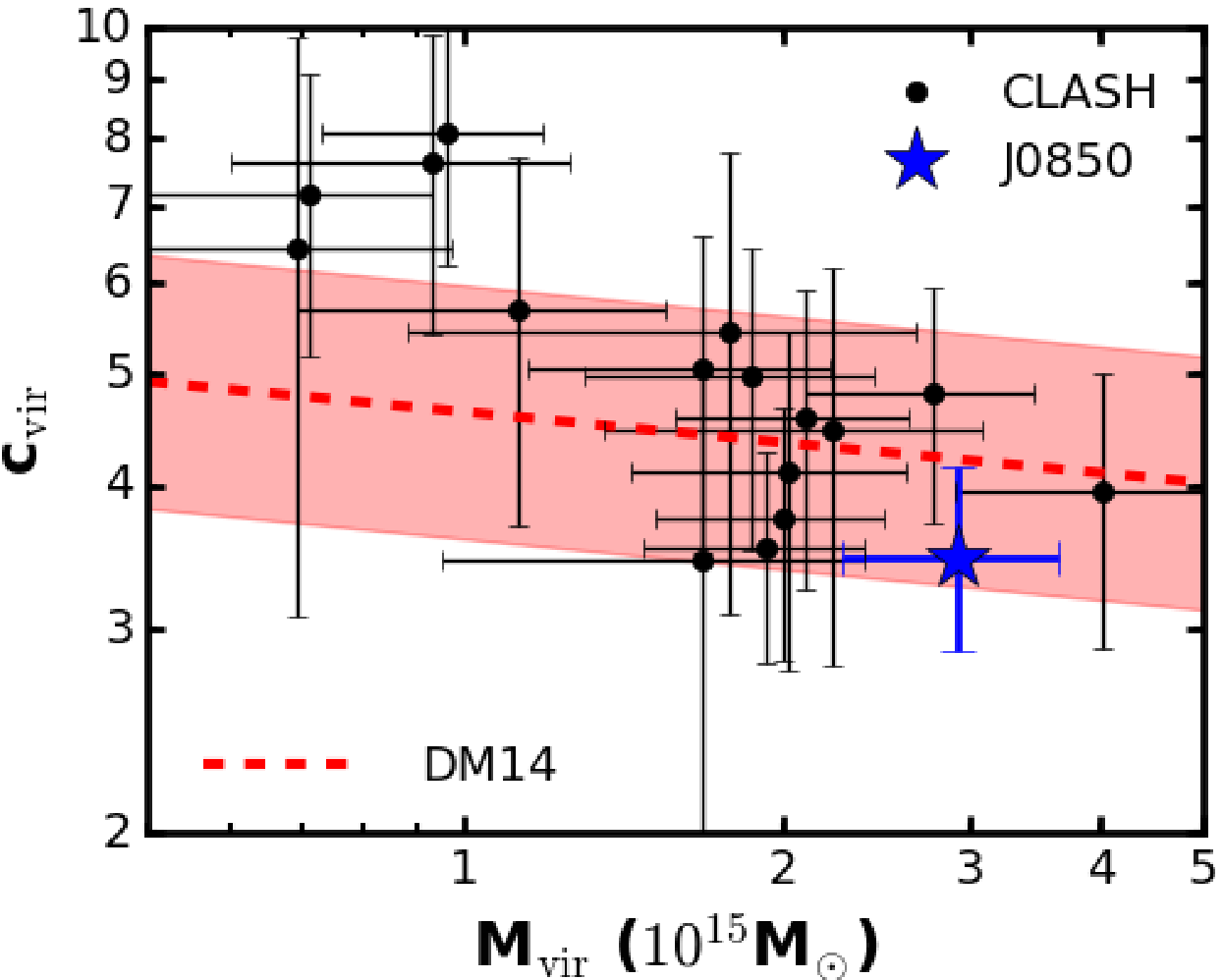}
\caption{$\mathrm{M_{vir}-c_{vir}}$ relation for massive clusters.  The main J0850 cluster (blue star) is plotted in comparison to the CLASH sample of X-ray-selected clusters \citep[black points;][]{umetsu2017}.  Also shown is the relation determined from simulations \citep[red line;][assuming the {\it Planck} cosmology]{dutton2014}.  The comparison samples are at similar masses and redshifts.  The cluster has a concentration consistent with both the CLASH and simulation results.  The uncertainties for the CLASH clusters are slightly larger than ours due to the inclusion of systematic errors arising from different model parameterizations.
\label{fig:mc_clash}}
\end{figure}

Compared to the results from weak lensing alone, the addition of strong lensing information from the single multiply-imaged source considerably tightens the constraints on the main halo's concentration by roughly a factor of two ($\mathrm{c_{vir}} = \csl$).  Past results from simulations and observations have shown that dark matter halos parameterized as NFW profiles show a negative correlation between halo mass and concentration \citep[e.g,][]{oguri2012,meneghetti2014,diemer2015,okabe2016,umetsu2016,umetsu2017} as a consequence of hierarchical structure formation.  It is also known that lensing-selected clusters can show a bias toward higher concentrations at a given mass due to their triaxial shapes and projection effects \citep[e.g.,][]{corless2007,hennawi2007,broadhurst2008,oguri2009,oguri2012,wong2012}.  In Figure~\ref{fig:mc_clash}, we show the virial mass and concentration of the main J0850 cluster in comparison to theoretical and observational results from the literature for comparable mass and redshift ranges.  The cluster has a concentration consistent with those of the CLASH X-ray-selected clusters \citep{umetsu2017} and derived from simulations \citep{dutton2014} assuming the {\it Planck} cosmology\footnote{Our assumed cosmology is slightly different from the {\it Planck} cosmology, but this difference has a negligible effect on the mass-concentration relation in Figure~\ref{fig:mc_clash}.}.  This field was selected by LRG luminosity density rather than by lensing, so the cluster should not suffer from an overconcentration bias due to projection effects,  Thus, it is not surprising to find a relatively typical concentration for its mass.

The addition of the strong lensing constraint shifts the cluster centroid $\sim17\arcsec$ in the direction of the \citet{ammons2014} centroid, which was determined from the average position of the spectroscopically confirmed cluster members.  The likely explanation is that the weak lensing alone provides a noisy centroid measurement, as the central regions of the cluster are ignored (Section~\ref{subsubsec:wlselect}).  The addition of the strong lensing constraint adds information from these central regions, allowing a more accurate measure of the mass centroid, as can be seen in the smaller uncertainty.  It is likely that constraints from additional multiply-imaged systems, particularly those at different locations along the tangential critical curve, will greatly improve the centroid determination.

The additional constraints provided by the strong lensing information in this analysis are coming from just a single pair of lensed images.  With deeper, higher-resolution imaging data, we would detect more strongly-lensed systems and achieve constraints similar to the CLASH or the HFF analyses \citep{johnson2016}, where many more lensed systems have been identified from {\it HST} imaging.  Yet, even with a single lensed image pair, we add meaningful constraints to the properties of the main J0850 cluster, a result with implications for the lensing fields identified in wide-area photometric surveys such as the Large Synoptic Survey Telescope (LSST).  Follow-up space-based observations of large numbers of clusters will be prohibitively expensive, and it may be necessary to rely on a limited number of constraints from ground-based data alone, as we have done here.

\begin{figure*}
\plotone{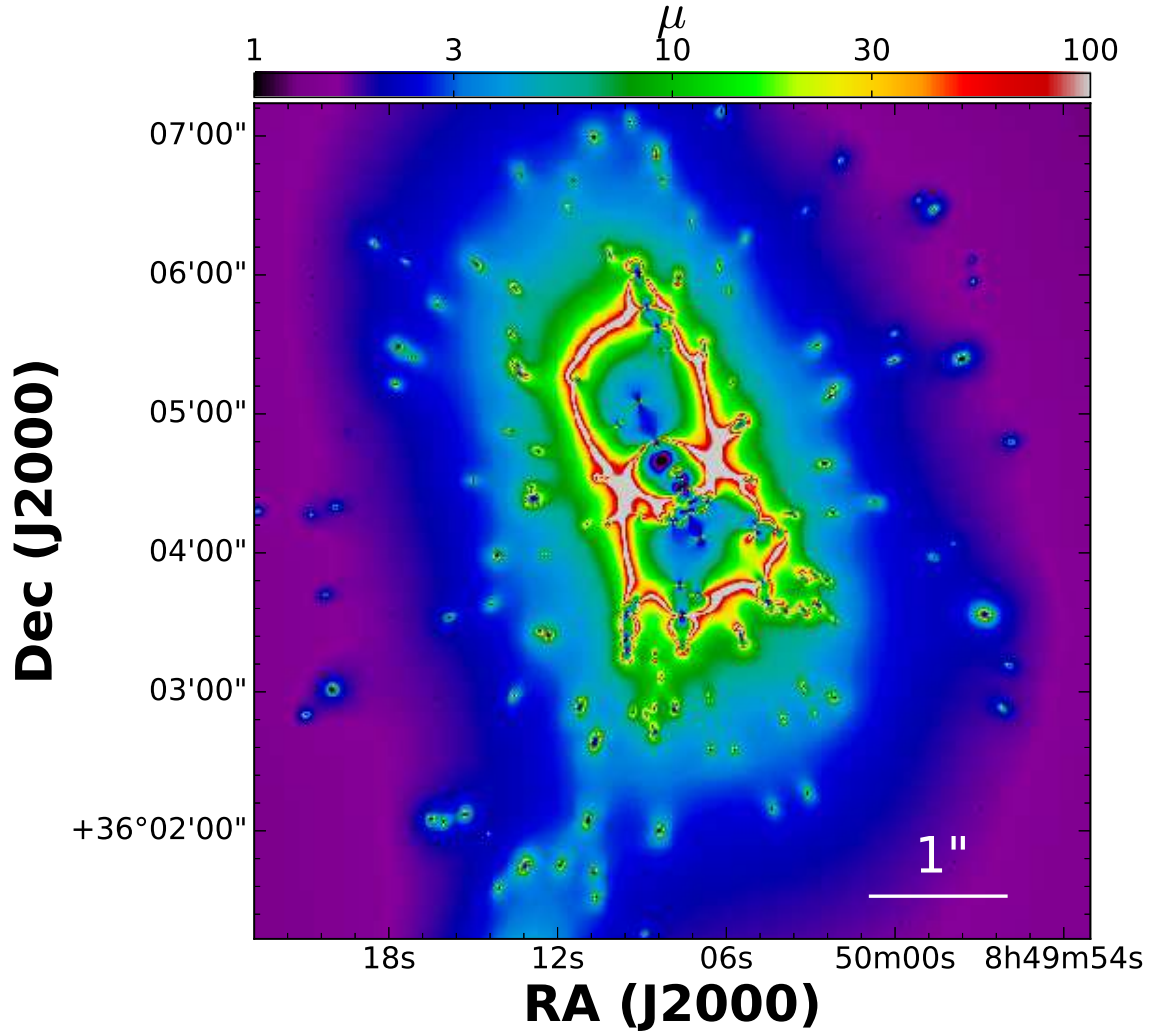}
\caption{Magnification map for the best mass model of the J0850 field for a source redshift of $z_{\mathrm{S}} = 5.03$ based on our combined lensing analysis.  The region shown is $6\arcmin \times 6\arcmin$, and the angular scale is indicated by the white bar in the bottom right corner.  The color bar indicates the magnification on a logarithmic scale.  The magnification map shows a large area of intermediate-to-high magnification, suggesting that this field is an excellent candidate to search for highly-magnified high-redshift galaxies.  The small influence of the foreground group can be seen in the southeastern part of the magnification map, but it does not affect the majority of the high-magnification region.
\label{fig:magmap}}
\end{figure*}

Figure~\ref{fig:magmap} is a magnification map of the J0850 field for a source redshift of $z_{\mathrm{S}} = 5.03$ using the best model from our combined strong and weak lensing analysis.  This best model shows characteristics similar to that of the ensemble distribution, such as a large ellipticity and a position angle slightly East of North.  The small influence of the foreground halo can be seen in the southeastern part of the magnification map, but it does not affect the majority of the high-magnification region.  The large area of intermediate-to-high magnification further suggests that this field is an excellent candidate to search for highly-magnified high-redshift galaxies.

\begin{figure*}
\plotone{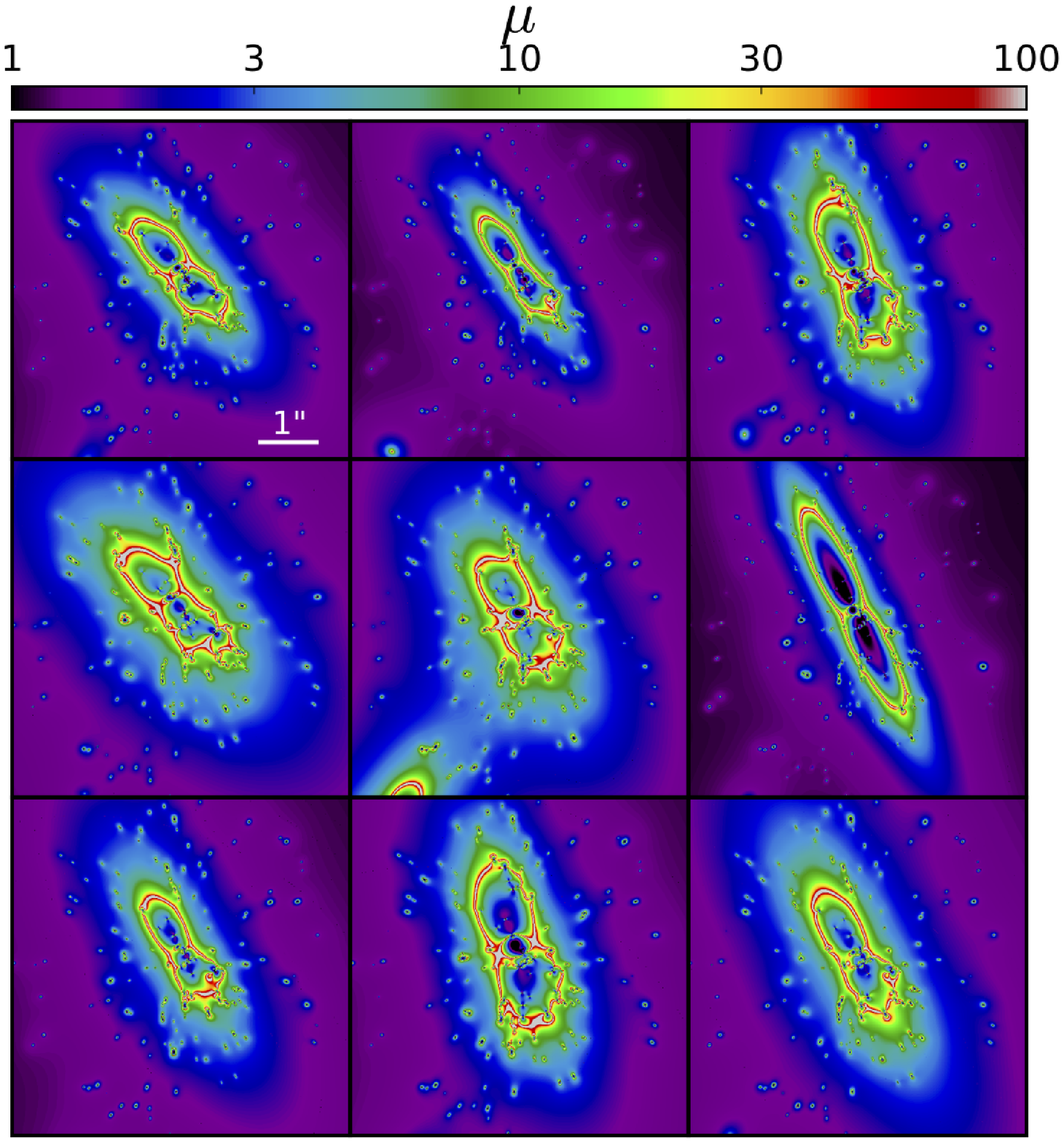}
\caption{Magnification map for a subsample of nine other successful models (that are able to reproduce both arcs) of the J0850 field for a source redshift of $z_{\mathrm{S}} = 5.03$.  The regions shown in each panel are the same $6\arcmin \times 6\arcmin$ area shown in Figure~\ref{fig:magmap}, and the angular scale is indicated by the white bar in the bottom right corner of the first panel.  The color bar indicates the magnification on a logarithmic scale.  Despite the diversity, the successful models have the same general shape and a large area of intermediate-to-high magnification.
\label{fig:magrange}}
\end{figure*}

To give a sense of the uncertainty in the magnification maps, we show similar maps for several other successful models in our sample in Figure~\ref{fig:magrange}.  Despite the diversity, the successful models have the same general shape and a large area of intermediate-to-high magnification.

\subsection{Highly-magnified Fold Arc} \label{subsec:fold_arc}
As noted in Section~\ref{subsubsec:slapp}, the majority of our successful strong lens models have the northern image of the lensed $z \approx 5.03$ background galaxy lying across a lensing caustic (Figure~\ref{fig:slmodel}).  This configuration is quite different from the typical method of producing a close pair of lensed images in which the source lies near but not across a lensing caustic and each image is of the entire source.  The part of the source that lies inside the lensing caustic has an extremely large magnification: the overall magnification for the northern arc is $\mu = 118$ for our best model, and the factor is even higher ($\mu > 1000$) for portions of the image close to the critical curve.  Past observations of cluster lenses have revealed arcs with extreme magnifications near the critical curves, but these have almost always been at lower redshifts since higher redshift objects are often intrinsically smaller.  Our model predicts that J0850 is a unique field in which a source is both at high redshift and is highly magnified, which is very rare.

Our model indicates that the small ($\sim 0\farcs1$) region of the source that lies within the caustic is elongated into a $\sim 1\farcs75$ image, potentially allowing us to study features at physical scales of $< 30$ pc with sufficiently high-resolution observations.  Thus, this unique configuration could provide new insights into resolved star formation and stellar populations at $z \approx 5$.

\section{Conclusions} \label{sec:conclusion}
Using ground-based imaging and spectroscopy, we have performed a joint weak and strong lensing analysis of the J0850 field containing the massive cluster Zwicky 1953 at $z = \zl$.  We present a new technique using multi-plane lensing effects to constrain the properties of the massive cluster while simultaneously accounting for a foreground structure at $z = \zfg$ and individual galaxies along the line of sight.  Unlike past studies, this technique accounts for the full three-dimensional mass structure along the line of sight.  Our methodology can be generalized to lines of sight containing multiple cluster-scale halos at distinct redshifts.  Other fields from the \citet{wong2013} sample have been confirmed to contain massive clusters with such configurations, making them new fields with large magnifications over much of the source plane with which to study the most distant and faint galaxies, complementary to the HFFs.

We confirm that J0850 contains one of the most massive known clusters, Zwicky 1953, with a virial mass of $\mathrm{M_{vir}} = \mclsl$ and a concentration of $\mathrm{c_{vir}} = \csl$, consistent with that of other clusters of similar mass and redshift.  Since this cluster was selected by its integrated LRG luminosity, it is not biased toward higher concentrations as purely lensing-selected clusters are.  The cluster is highly elliptical, suggesting that it has a high lensing efficiency, and its large mass makes it an ideal cosmic telescope for studying background sources.  Despite having only a single multiply-imaged galaxy from our ground-based imaging, we are able to tighten the constraints on the halo concentration by a factor of two, as well as marginally improve the constraints on the halo centroid and mass compared to the weak lensing analysis alone.  This result highlights the importance of complementary strong lensing constraints, even just from a single pair of images.

Using the full surface brightness distribution of the pair of lensed images (rather than only their positions) is highly constraining because of the way it pins down the location of the lensing critical curve.  In fact, our models predict that the source galaxy crosses a lensing caustic such that the northern arc is actually a merged pair of images from the galaxy's outskirts.  This fold arc is highly magnified, which means that high-resolution imaging would have the potential to resolve structure in the source galaxy on very small ($< 30$ pc) scales.  Thus, J0850 is a unique field with a source that is both at high redshift and is highly magnified, offering a chance to study detailed properties of a galaxy at $z \approx 5$.

We determine a magnification map for this field based on our lens models, which will be used to search for high-redshift sources and constrain their intrinsic properties.  As we move into the era of the Large Synoptic Survey Telescope, which will identify many more lensing fields, the ability to determine magnification maps without expensive, follow-up space-based deep imaging will be critical.  Our analysis demonstrates a self-consistent methodology to analyze multi-plane lensing fields using a combination of weak and strong lensing using only ground-based observations, paving the way forward for taking advantage of newly-discovered cosmic telescopes to access the early universe.

\acknowledgments
We thank the referee for helpful comments, which improved this paper.  We thank Olivier Guyon and Curtis McCully for their contributions to this project.  We thank Amit Tagore for assisting with the {\sc pixsrc} software.  We thank Tom Broadhurst for useful discussions and input.
This work is based in part on data collected at Subaru Telescope and obtained from the SMOKA, which is operated by the Astronomy Data Center, National Astronomical Observatory of Japan.
K.C.W. is supported by an EACOA Fellowship awarded by the East Asia Core Observatories Association, which consists of the Academia Sinica Institute of Astronomy and Astrophysics, the National Astronomical Observatory of Japan, the National Astronomical Observatories of the Chinese Academy of Sciences, and the Korea Astronomy and Space Science Institute.
C.R.\ and C.R.K.\ acknowledge support from NSF grant AST-1211385.
K.U. acknowledges support from the Ministry of Science and Technology of Taiwan (grants MOST 103-2112-M-001-030-MY3 and MOST 103-2112-M-001-003-MY3).
A.I.Z. acknowledges NSF grant AST-1211874.
Portions of this work were performed under the auspices of the U.S. Department of Energy by Lawrence Livermore National Laboratory under Contract DE-AC52-07NA27344.

\end{document}